# Improving the doping efficiency of Al in 4H-SiC by co-doping group-IVB elements

Yuanchao Huang[1,2], Rong Wang,[2*] Yixiao Qian[2], Yiqiang Zhang[3], Deren Yang[1,2], Xiaodong Pi[1,2*]

[1]*State Key Laboratory of Silicon Materials and School of Materials Science and Engineering, Zhejiang University, Hangzhou, 310027, China*

[2]*Hangzhou Global Scientific and Technological Innovation Center, Zhejiang University, Hangzhou, 311200, China*

[3]*School of Materials Science and Engineering & Henan Institute of Advanced Technology, Zhengzhou University, Zhengzhou, Henan 450001, China*

**ABSTRACT**

The *p*-type doping efficiency of 4H silicon carbide (4H-SiC) is rather low due to the large ionization energies of *p*-type dopants. Such an issue impedes the exploration of the full advantage of 4H-SiC for semiconductor devices. In this letter, we show that co-doping group-IVB elements effectively decreases the ionization energy of the most widely used *p*-type dopant, i. e., aluminum (Al), through the Coulomb repulsion between the energy levels of group-IVB elements and that of Al in 4H-SiC. Among group-IVB elements Ti has the most prominent effectiveness. Ti decreases the ionization energy of Al by nearly 50%, leading to a value as low as ~ 0.13 eV. As a result, the ionization rate of Al with Ti co-doping is up to ~ 5 times larger than that without co-doping at room temperature when the doping concentration is up to $10^{18}$ cm$^{-3}$. This work may encourage the experimental co-doping of group-IB elements such as Ti and Al to significantly improve the *p*-type doping efficiency of 4H-SiC.

[*]Corresponding authors.

E-mail address: rong_wang@zju.edu.cn; xdpi@zju.edu.cn

As a wide bandgap semiconductor, silicon carbide (SiC) has been known as a good material for semiconductor devices since early the 20th century.[1-3] Among all kinds of SiC polymorphs 4H-SiC has attracted great attention given its superior electrical properties.[1] The effective *n*-type doping of 4H-SiC may be readily carried out by doping nitrogen (N) because the ionization energy of N is rather small (~ 0.06 eV).[4-6] However, the efficiency of the *p*-type doping of 4H-SiC has remained poor despite its long-time development. This is mainly attributed to the large ionization energies of *p*-type dopants in 4H-SiC. For example, the most widely used *p*-type dopant of aluminum (Al) has the ionization energy of ~ 0.23 eV, which leads to the ionization rate of no more than ~ 30% for the typical doping concentration of Al in the range from $10^{16}$ to $10^{18}$ cm$^{-3}$ at room temperature.[7-11] The low ionization rate of Al means that large amount of unionized Al is in 4H-SiC, limiting the mobility of holes by impurity scattering.[9] In addition, the carrier-capture/emission of unionized Al poses severe reliability issues for devices based on 4H-SiC.[8] Hence, the poor doping efficiency of *p*-type dopants in 4H-SiC needs to be well addressed especially given the fact that the development of 4H-SiC has recently gained great momentum for power electronics and radio-frequency electronics.[1-3] A series of important 4H-SiC devices such as *p*-type SiC metal-oxide semiconductor field effect transistors (MOSFETs)[12] and *n*-channel SiC insulated gate bipolar transistors (IGBTs)[13-15] demand that 4H-SiC should be much more effectively doped with *p*-type dopants.

Various approaches such as host alloying[16-19], strain modulation[20], incorporation of band-like defect-induced energy levels [21] and co-doping[22-30] have been proposed to lower the ionization energies of dopants in wide-bandgap semiconductors. Among these approaches co-doping is attractive due to its simple process and negligible damage to the host. When a *p*- or *n*-type dopant is co-doped with other dopants, Coulomb repulsion may push the energy level of the *p*- or *n*-type dopant to a shallower position, decreasing the ionization energy of the *p*- or *n*-type dopant. The effectiveness of co-doping strongly depends on the symmetries of the energy levels of dopants. By taking advantage of co-doping researchers have successfully improved the efficiency of the *n*-type doping of diamond[22-24] and ZnTe[25] and the *p*-type doping of ZnO[26-30].

In this letter, we demonstrate that co-doping is also an effective means to improve the doping efficiency of the most widely used *p*-type dopant, *i.e.*, Al, in 4H-SiC. By analyzing the symmetry of the energy levels of elements throughout the whole periodic table in the framework of crystal field theory, we first identify that the energy levels of group-IVB elements have the same symmetry as that of Al. The coupling between the energy levels of group-IVB elements and that of Al should push the energy level of Al towards the valance band maximum (VBM), reducing the ionization energy of Al. Subsequent first-principles calculations verify that the unoccupied energy levels of group-IVB elements indeed interact with the acceptor level of Al, effectively lowering the ionization energy of Al. We find that Ti has the most prominent effectiveness among all the group-IVB elements owing to its lowest orbital energies and smallest atomic size. The co-doping of Ti decreases the ionization energy of Al by nearly 50%, leading to a value as low as ~ 0.13 eV. As a result, the ionization rate of Al with Ti co-doping may be up to ~ 5 times larger than that without co-doping at room temperature when the concentration of Al is up to $10^{18}$ cm$^{-3}$.

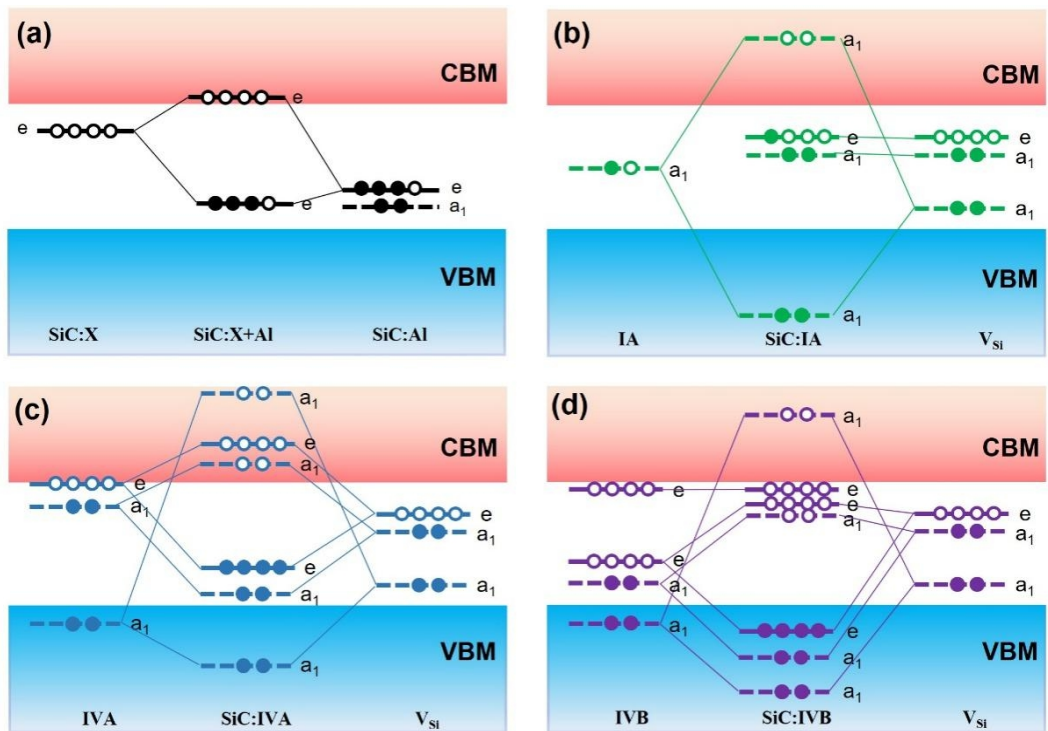

Figure 1 (a) Schematic illustration of the Coulomb repulsion between the energy levels of Al and an ideal impurity $X$ with an upper empty $e$ state. The symmetry analysis for the energy levels of (b) group-IA, (c) group-IVA and (d) group-IVB elements is carried out based on crystal field theory.

It is well known that Al prefers substituting Si in 4H-SiC.[11] As shown in Figure 1(a), the orbital of Al splits into a fully occupied single $a_1$ state and a 3/4 occupied double-degenerated $e$ state in the $C_{3v}$ symmetry of 4H-SiC. The $e$ state can capture an electron from the VBM of 4H-SiC, enabling the $p$-type doping of Al. In an ideal co-doping case, an impurity $X$ should introduce an upper unoccupied $e$ state in the bandgap of 4H-SiC. The coupling between the $e$ states of $X$ and Al pushes the $e$ state of Al to a shallower position, improving the efficiency of the $p$-type doping of Al. According to crystal field theory,[31] for the $C_{3v}$ symmetry of 4H-SiC a $s$ orbital splits into one $a_1$ state. A $p$ orbital splits into one $a_1$ state and one $e$ state. A $d$ orbital splits into one $a_1$ state and two $e$ states. When $X$ substitutes Si, the orbitals of $X$ couple with those of a Si vacancy ($V_{Si}$), introducing the states of $X$ in 4H-SiC. Due to the different arrangements of valence electrons of elements throughout the periodic table, $X$ should have quite different states in 4H-SiC. We classify $X$ into three groups, which are the $s$ group, $sp$ group and $ds$ group.

The $s$ group elements contain group-IA and group-IIA elements. Their valence electrons occupy the $s$ orbital. Taking group-IA elements as an example, we can see that the $s$ orbital splits into one $a_1$ state. $V_{Si}$ has two fully occupied $a_1$ states and an empty $e$ state. The $a_1$ state of an element in the $s$ group may interact with the $a_1$ states of $V_{Si}$, forming a fully occupied $a_1$ state and an empty $a_1$ state [Fig. 1(b)]. The electron occupying the $a_1$ state of the $s$-group element transfers to the unoccupied $e$ state of $V_{Si}$. This means that an element in the $s$ group cannot introduce an ideally unoccupied $e$ state in the bandgap. Therefore, $s$-group elements may not be used as co-dopants for Al in 4H-SiC.

The $sp$ group elements contain group-IIIA, IVA, VA, VIA and VIIA elements. Elements in group-VA, VIA and VIIA have more than four valence electrons, offering extra electrons to compensate the 3/4 occupied $e$ state of Al. Hence, they are not suitable to be co-dopants for Al. For group-IVA elements they have two fully occupied $a_1$ states and an occupied $e$ state near the VBM and two fully unoccupied $a_1$ states and an unoccupied $e$ state inside the conduction band, as shown in Fig 1(c). Since the $p$ orbital energy of Ge/Sn/Pb in group-IVA is higher than that of Si, the formed occupied states of Ge/Sn/Pb are higher than the VBM. The formed unoccupied states of Ge/Sn/Pb are higher than the CBM. This occupied $e$ state would repel the $e$ state of Al toward a higher energy position in the bandgap, increasing the ionization energy of Al. For group-IIIA elements, because their valence electrons are one fewer than Si, the 3/4 occupied $e$ state of a group-IIIA element will be pushed into the bandgap by the host.[32] After co-doping a group-IIIA element with Al, Coulomb repulsion may occur between the two 3/4 occupied $e$ state, forming a lower fully occupied $e$ state and a higher 2/4 occupied $e$ state. This leads to a higher ionization energy. Therefore, elements in the $sp$ group cannot be ideally used as co-dopants for Al, either.

Elements in the $ds$ group are those in groups from IB to VIIIB. We may readily exclude elements in groups from VB to VIIIB because they can introduce extra electrons to compensate the $e$ state of Al. As for group IB, IIB, and IIIB elements, owing to fewer valence electrons compared with Si they introduce partially occupied $e$ states in the bandgap, which further couple with the 3/4 $e$ state of Al to form a higher acceptor level. The energy levels of a group-IVB element are schematically shown in Fig. 1(d). The valence electrons of group-IVB elements have the $d^2s^2$ arrangement. In the $C_{3v}$ symmetry of 4H-SiC a $s$ orbital splits into one full occupied $a_1$ state. A $d$ orbital splits into one full occupied $a_1$ state and two unoccupied $e$ states. Because the $d$ orbital energy is lower than that of the $p$ orbital of Si,[33] after interacting with the states of $V_{Si}$, the $d$ orbital gives rise to occupied $e/a_1$ states that are lower than the VBM and unoccupied $a_1/e/e$ states that are lower than the CBM. Therefore, group-IVB elements have the potential to lower the $e$ state of Al by using the Coulomb repulsion between the $e$ states of group-IVB elements and Al, effectively decreasing the ionization energy of Al.

Now we carry out first-principles calculations to examine the effect of the co-doping of group-IVB elements such as Ti, Zr and Hf on the ionization energy of Al in 4H-SiC. The first-principles calculations are performed by using the Vienna *ab-initio* Simulation Package (VASP). The generalized gradient approximation (GGA) in with the Perdew, Burke, and Ernzerhof (PBE) functionals is employed to describe the exchange-correlation interactions.[34] A 4 × 4 × 1 supercell of 4H-SiC with 128 atoms is constructed. A single-doping model is constructed by replacing one Si atom with one dopant atom, while a co-doping model is built by replacing two neighboring Si atoms with two different dopant atoms. The plane-wave energy cutoff is 500 eV. The *k*-mesh is set to be 2 × 2 × 2. During the structural relaxations a conjugate-gradient algorithm is used. In a relaxed structure the force on each atom is lower than 0.01 eV/Å. The total energy is converged to $1.0 \times 10^{-6}$ eV. To accurately calculate the bandgap, we use static calculations with the hybrid functional of Heyd, Scusseria and Ernzherof (HSE06) to work on the relaxed structures obtained with the PBE functionals. Our calculation shows that the calculated lattice parameters of 4H-SiC are a = 3.07 Å and c = 10.05 Å. The calculated bandgap of 4H-SiC is 3.18 eV. These are well consistent with experimental results.[1]

The formation energy of a dopant atom at the charge state of *q* in 4H-SiC is defined as[35,36]

$$\Delta H_f(dopant, q) = \Delta E(dopant, q) + \sum n_i \mu_i + qE_F, (1)$$

where $\Delta E(dopant, q) = E(dopant, q) - E(host) + \sum n_i E(i) + q\varepsilon_{VBM}$. $E(dopant, q)$ is the energy of a supercell containing the dopant atom at the charge state of *q*. $E(host)$ is the energy of the host in the same supercell without the dopant. $\varepsilon_{VBM}$ is the energy of the VBM of the host. $E_F$ is the Fermi energy referenced to the VBM. $\mu_i$ is the chemical potential of constituent *i* referenced to elemental solid or gas with energy $E(i)$. $n_i$ is the number of atoms removed from or added into the supercell. In this work, two chemical-potential conditions (C-rich and Si-rich) are considered due to the limitation of the formation energy of bulk SiC. The charge transition energy level with respect to the VBM is calculated in the mixed *k*-point scheme by

using[36]

$$\varepsilon(0/q) = \left[\varepsilon_D^{\Gamma}(0) - \varepsilon_{VBM}^{\Gamma}(host)\right] + \left\{E(dopant, q) - \left[E(dopant, 0) - q\varepsilon_D^{k}(0)\right]\right\}/(-q), (2)$$

where $\varepsilon_D^{\Gamma}(0)$ is the single-electron energy level of the impurity at the $\Gamma$ point, $\varepsilon_{VBM}^{\Gamma}(host)$ is the energy of the VBM at the $\Gamma$ point, and $\varepsilon_D^{k}(0)$ is the averaged impurity energies weighted over $k$ points. The first term on the right-hand side of Eq. (2) is the vertical activation energy, while its second term determines the structural relaxation energy after charging.

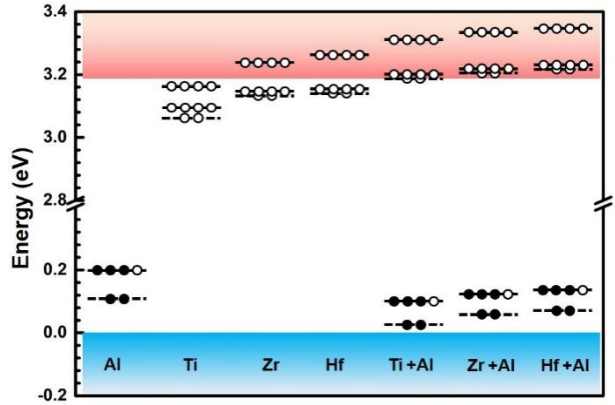

Figure 2 Single-electron levels of Al, Ti, Zr, Hf, Ti+Al, Zr+Al and Hf+Al in 4H-SiC. All the dopants substitute Si in 4H-SiC.

The calculated total energies suggest negligible difference (within 0.02 eV) between the substitutional $k$ and $h$ sites of Si for Al, Ti, Zr and Hf. Therefore, we only consider that Al, Ti, Zr and Hf substitute $k$-site Si in this work. We first examine the single-electron levels of all the dopants, which actually correspond to the vertical activation energies in Eq. (2). The results are shown in Figure 2. We can see that the fully occupied $a_1$ state of Al is located at $E_v$ + 0.11 eV. The 3/4 occupied $e$ state of Al is located at $E_v$ + 0.21 eV. For Ti, Zr and Hf they each introduce an unoccupied $a_1$ state and two $e$ states near the CBM, which is consistent with aforementioned symmetry analysis. When $X$ changes from Ti to Hf, the unoccupied $a_1$ state and two $e$ states of $X$ shift upward because the $d$ orbital energy increases. Among group-IVB elements the states of Ti are the lowest due to its lowest $d$ orbital energy and smallest atomic size. After co-doping Al with Ti, Zr and Hf we observe that the 3/4 occupied $e$ state of Al shifts to $E_v + 0.11$, $E_v + 0.13$ and $E_v + 0.15$ eV, respectively. The unoccupied $a_1$ state and two $e$ states of Ti/Zr/Hf near the CBM are pushed into the conduction band of 4H-SiC. These confirm the effective coupling between Al and group-IVB elements.

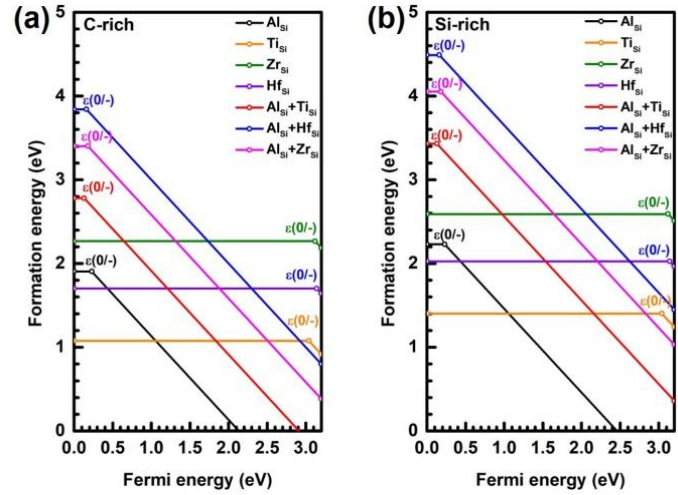

Figure 3 Formation energies of Al, Ti, Zr, Hf, Ti+Al, Zr+Al and Hf+Al at the (a) C-rich limit and (b) Si-rich limit.

We have then calculated the charge transition energies and formation energies of Al, Ti, Zr, Hf, Ti+Al, Zr+Al and Hf+Al, as shown in Figure 3. The calculated charge transition energies and formation energies of neutral dopants are tabulated in Table 1. The formation energy of neutral Al is 1.9 and 2.2 eV at the C-rich limit and Si-rich limit, respectively. The calculated (0/-) transition energy of Al, namely the ionization energy, is $E_V+0.23$ eV, which is in agreement with experiment results.[1] For Ti, Zr, and Hf the formation energies are 1.08/1.40, 2.27/2.59 and 1.70/2.03 eV at the C-rich/Si-rich limit, respectively. Their (0/-) transition energies are $E_C-0.17$, $E_C-0.08$, and $E_C-0.06$ eV, respectively. Please note that the current results for Ti are well consistent with experimental observation. It was found that Ti in the graphite parts of a 4H-SiC single-crystal growth equipment might be unintentionally incorporated into 4H-SiC,[37,38] indicating the small formation energy of Ti. Dalibor et al. once observed a very deep acceptor level at $E_C-0.17$ eV in 4H-SiC by using deep-level transient spectroscopy,[37] exactly agreeing with the currently obtained (0/-) transition energy at $E_C-0.17$ eV for Ti. With the co-doping of Ti, Zr and Hf the ionization energy of Al significantly decreases to $E_V+0.13$, $E_V+0.16$ and $E_V+0.18$ eV, respectively. It is clear that the co-doping of Ti has the most prominent effectiveness, which decreases the ionization energy of Al by nearly 50%. This is due to the lowest *d* orbital energy of Ti, which induces the strongest Coulomb repulsion with Al. In the meantime, Ti has the smallest atomic size, leading to the smallest structural relaxation energy.

The stability of Ti+Al, Zr+Al and Hf+Al is subsequently checked by calculating their binding energies with $E_b = \Delta H_f(IVB) + \Delta H_f(Al) - \Delta H_f(IVB + Al)$. The results are shown in Table 1. The values of $E_b$ for Ti/Zr/Hf+Al are all positive, indicating that the formation of Ti/Zr/Hf+Al is favorable in energy when Ti/Zr/Hf and Al coexist. Ti/Zr/Hf can form stable complexes with Al, effectively doping 4H-SiC.

Table 1 Charge transition energies, formation energies of neutral defects and binging energies

| Defects | Charge transition energy (eV) | Formation energy of neutral defect (eV) | | Binding energy of X+Al (eV) |
|---|---|---|---|---|
| | | C-rich | Si-rich | |
| Al | $E_V+0.23$ | 1.90 | 2.23 | - |
| Ti | $E_C-0.17$ | 1.08 | 1.40 | - |
| Z | $E_C-0.08$ | 2.27 | 2.59 | - |
| Hf | $E_C-0.06$ | 1.70 | 2.03 | - |
| Ti+Al | $E_V+0.13$ | 2.78 | 3.43 | 0.20 |
| Zr+Al | $E_V+0.16$ | 3.84 | 4.49 | 0.33 |
| Hf+Al | $E_V+0.18$ | 3.40 | 4.05 | 0.20 |

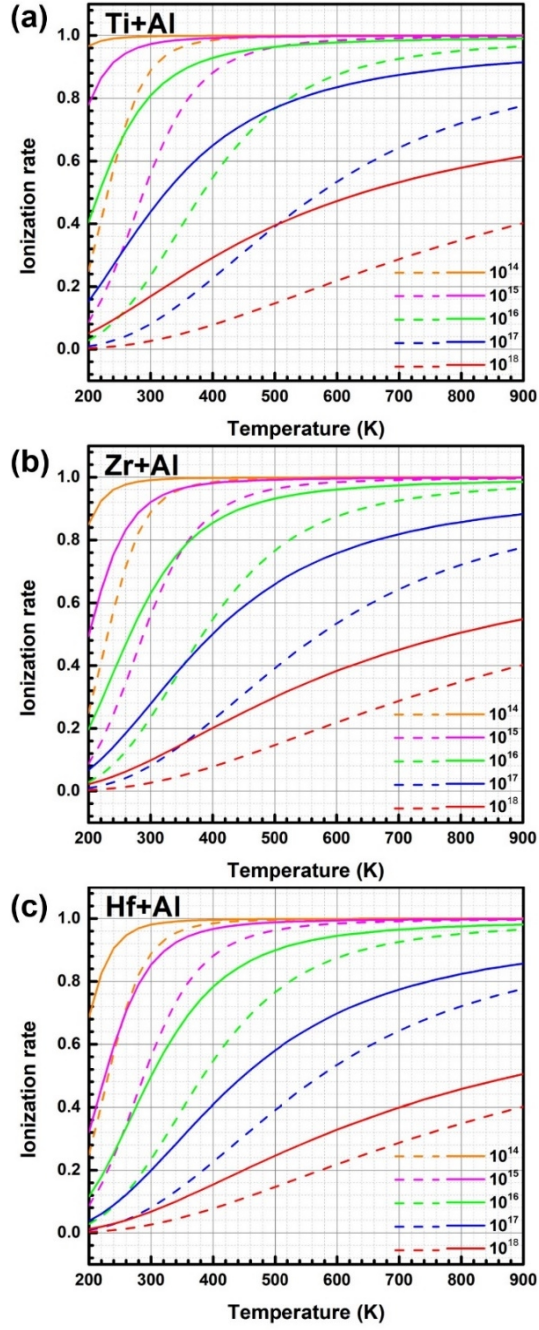

Figure 4 The ionization rate of Al with the co-doping of (a) Ti, (b) Zr, and (c) Hf at different doping concentrations as a function of temperature. The dashed line and solid lines represent the ionization rate of Al and Ti/Zr/Hf+Al, respectively.

We finally investigate the ionization rate of Al through calculating the concentration of ionized Al ($N_A^-$) at thermodynamic equilibrium by using[1]

$$N_A^- = \frac{\eta}{2}\left(\sqrt{1 + \frac{4N_A}{\eta}} - 1\right), (3)$$

where $\eta$ is given by $\eta = N_V/g_A \exp(-\varepsilon(0/-)/kT)$, $\varepsilon(0/-)$ is the ionization energy, $g_A$ is the degeneracy factor of Al, and $N_V$ is the effective density of states in the valence band given by $N_V = 2(2\pi m_{dh}^* kT/h^2)^{3/2}$. Please note that $m_{dh}^*$ is the density-of-states effective mass for holes, $h$ is the Planck's constant, $k$ is the Boltzmann constant, and $T$ is the temperature. As shown in Figure 4, the ionization rate is found to decrease with the increase of the concentration of Al in the typical range from $10^{14}$ to $10^{18}$ cm$^{-3}$. When only Al is doped, the ionization rate of Al at room temperature is ~ 88%, 56%, 24%, 8% and 3% for the concentrations of $10^{14}$, $10^{15}$, $10^{16}$, $10^{17}$ and $10^{18}$ cm$^{-3}$, respectively. With the co-doping of Ti/Zr/Hf the ionization rate of Al is significantly improved. Among these co-dopants the improvement induced by Ti is the most pronounced. For the Al concentrations of $10^{14}$, $10^{15}$, $10^{16}$, $10^{17}$ and $10^{18}$ cm$^{-3}$ the co-doping of Ti makes the ionization rate of Al reach 100%, 96%, 80%, 44% and 16% at room temperature, respectively. Clearly, for the highest concentration of $10^{18}$ cm$^{-3}$ the Ti-induced improvement of the ionization rate of Al is the most significant (a factor of ~ 5).

In summary, we have found that group-IVB elements may be employed to co-dope Al, effectively lowering the ionization energy of Al via the coupling between the energy levels of group-IVB elements and that of Al. Among the group-IVB elements Ti the most significantly reduces the ionization energy of Al (~ 50%), leading to a value of ~ 0.13 eV. Such a low ionization energy enables the ionization rate of Al to be up to ~ 5 times larger than that without co-doping at room temperature when the concentration of Al is up to $10^{18}$ cm$^{-3}$. Given the fact that group-IVB elements such as Ti have already been incorporated into 4H-SiC during its growth,[37,38] the current work should inspire the intentional doping of group-IVB elements together with Al during the materials preparation of 4H-SiC to effectively improve the doping efficiency of Al.

## ACKNOWLEDGEMENTS

This work is supported by National Key Research and Development Program of China

(Grant Nos. 2017YFA0205704, 2018YFB2200101), Natural Science Foundation of China (Grant Nos. 91964107, 61774133), Fundamental Research Funds for the Central Universities (Grant No. 2018XZZX003-02), Natural Science Foundation of China for Innovative Research Groups (Grant No. 61721005) and Zhejiang University Education Foundation Global Partnership Fund. National Supercomputer Center in Tianjin is acknowledged for computational support.

## DATA AVAILABILITY

The data that support the findings of this study are available from the corresponding author upon reasonable request.